\documentclass[aps,nofootinbib,superscriptaddress]{revtex4}
\usepackage{epsfig}
\newcommand{\be}{\begin{equation}}
\newcommand{\ee}{\end{equation}}

\begin{document}
\title{Modelling Nuclear Effects in Neutrino Scattering}
\thanks{Prepared for the International School of Nuclear Physics: Neutrinos in Cosmology, in Astro, Particle and Nuclear Physics, Erice, Italy, Sept. 16-24, 2005}
\author{T.~Leitner}
\email[corresponding author: ]{tina.j.leitner@theo.physik.uni-giessen.de}
\affiliation{Institut f\"ur Theoretische Physik, Universit\"at
Giessen, Germany}

\author{L.~Alvarez-Ruso}\affiliation{Institut f\"ur Theoretische Physik, Universit\"at
Giessen, Germany}
\author{U.~Mosel}
\affiliation{Institut f\"ur Theoretische Physik, Universit\"at
Giessen, Germany}

\date{November 21, 2005}

\begin{abstract} 
We have developed a model to describe the interactions of neutrinos with nucleons and nuclei via charged and neutral currents, focusing on the region of the quasielastic and $\Delta (1232)$ peaks. For $\nu N$ collisions a fully relativistic formalism is used. The extension to finite nuclei has been done in the framework of a coupled-channel BUU transport model where we have studied exclusive channels taking into account in-medium effects and final state interactions.

\end{abstract}

\maketitle

%\section{Introduction}
The detailed theoretical understanding of the weak nuclear response is a prerequisite for the analyses of neutrino experiments and a precise knowledge of the neutrino nucleus cross section is therefore essential. This situation motivates us to study neutrino scattering on nucleons and nuclei at low and intermediate energies up to about 2~GeV.
For a full description of our model we refer to \cite{leitner}.

%\section{Model}
In the impulse approximation, the neutrino nucleus interaction can be treated as a two-step process. The first step is the interaction of the neutrino with a single bound nucleon in the target nucleus. Then, as second step, the produced particles are propagated through the nucleus using the BUU transport model including a full coupled-channel treatment of the final state interactions (FSI). 

The nucleons are distributed in coordinate space according to Woods-Saxon and, in momentum space, following a local density approximation with local Fermi momenta.
The nucleons are bound in a momentum and density dependent mean field potential (cf.~\cite{teis}).

In the energy region under consideration, the elementary $\nu N$ reaction is dominated by two processes, namely quasielastic scattering and $\Delta$ production: 
For the charged current (CC) reaction, we consider
\be
\nu n \to l^- p, \quad \quad \nu n \to l^- \Delta^+, \quad \quad \nu p \to l^- \Delta^{++},\nonumber
\ee
which we describe with a fully relativistic formalism with state-of-the-art parametrizations of the form factors for both the nucleon and the $N-\Delta$ transition. For quasielastic scattering, we further take into account that the effective masses, \textit{i.~e.}~$M_{eff}=M + U(\vec{r},\vec{p})$, of in- and outgoing nucleons are not equal due to the mean field potential. In the nucleus, the produced nucleons might be Pauli blocked.
For $\Delta$ excitation, we additionally consider that the $\Delta$ is less bound in the nucleus than nucleons; we also include the in-medium modification of the $\Delta$ resonance due to Pauli blocking and collisional broadening:
\be
\Gamma^{med}_{tot} = \tilde{\Gamma} + \Gamma_{coll}.\nonumber
\ee
Further details on the elementary and the medium modified $\nu N$ reactions can be found in \cite{leitner,alvarez}.

Final state interactions are implemented by means of the coupled-channel Boltzmann-Uehling-Uhlenbeck (BUU) transport model which has been very successfully applied to $\gamma A$ and $eA$ scattering \cite{lehr}. It is based upon the BUU equation
\be 
\left({\partial_t}+\vec\nabla_p H\cdot\vec\nabla_r
  -\vec\nabla_r H\cdot\vec\nabla_p\right)F_i(\vec r,\vec p,\mu;t)
= I_{coll}[F_i,F_N,F_\pi,F_{\Delta},...], \nonumber
\ee
which describes the space-time evolution of the phase space density $F_i(\vec r,\vec p,\mu;t)$ of particles of type $i$ under the Hamiltonian $H$ in a mean field potential. The BUU equations are coupled via the mean field in $H$ and via $I_{coll}$. The collision integral $I_{coll}$ accounts for elastic and inelastic collisions, decays and the formation of resonances, including Pauli blocking. Those FSI therefore lead to absorption, charge exchange, a redistribution of energy and to the production of new particles (besides the nucleons, mesons and baryonic resonances are included). 
More precisely, in the initial $\nu N$ reaction both nucleons and $\Delta$'s are produced.  
The $\Delta$ might decay or interact via $\Delta N \to N N$, $\Delta N N \to N N N$, $\Delta N \to \pi N N$ or $\Delta N \to \Delta N$. The nucleons undergo elastic and charge exchange scattering with the nucleons in the medium via $NN \to NN$, $NN \to NN\pi$ or $NN \to N\Delta$. Thus, pions can be 
created or be absorbed in secondary collisions. 
For the pions, most important are $\pi N \to \pi N$, $NN\pi \to NN$ and $\pi N \to \Delta$, \textit{i.~e.}~elastic scattering with nucleons, charge exchange or absorption. 
In between the collisions, all particles are propagated (resonances are propagated off-shell) according to the BUU equation.

%\section{Results}

\begin{figure}[tb]  
\centering \epsfig{file=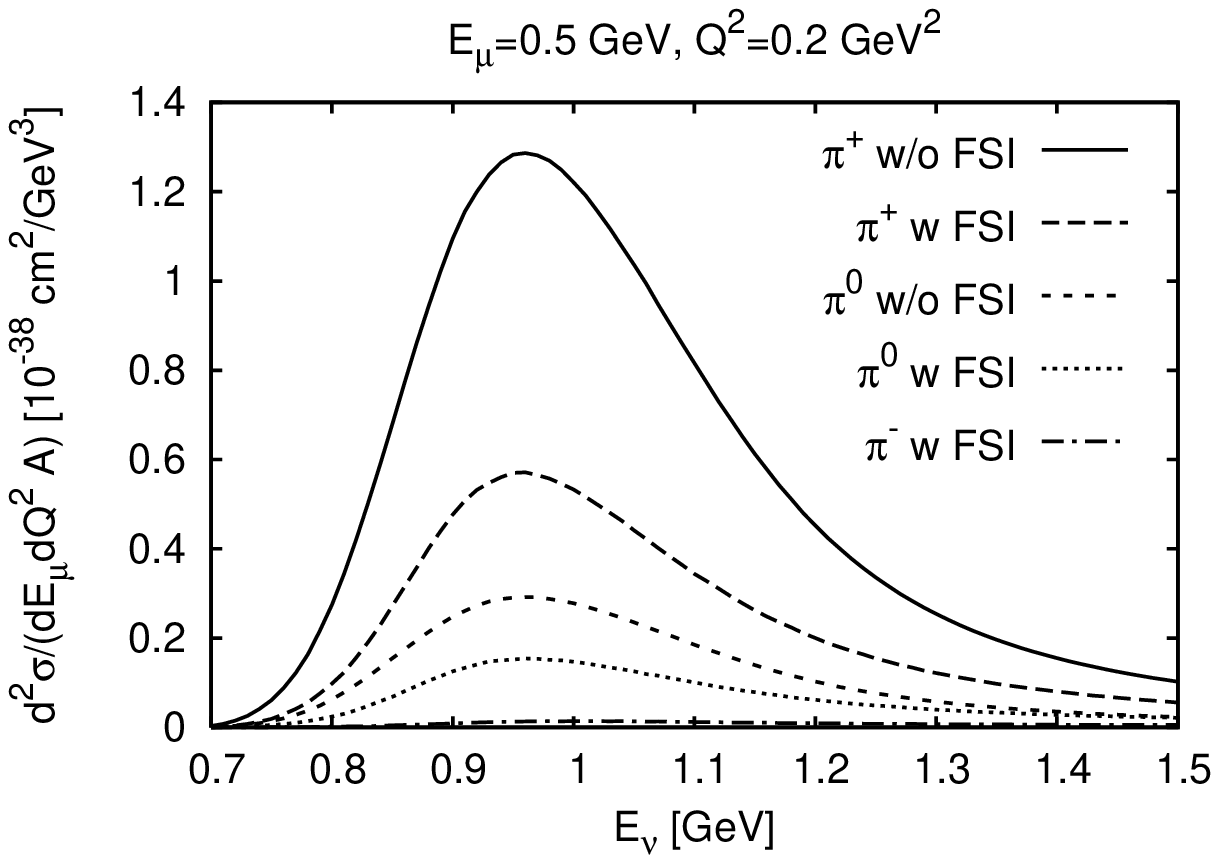,scale=.65}  
\caption{Exclusive cross section for $\nu_{\mu}$ induced $\pi^+$, $\pi^0$ and $\pi^-$ production on $^{56}$Fe. \label{fig:pionprod}}  
\vspace{0.5cm} 
\centering \epsfig{file=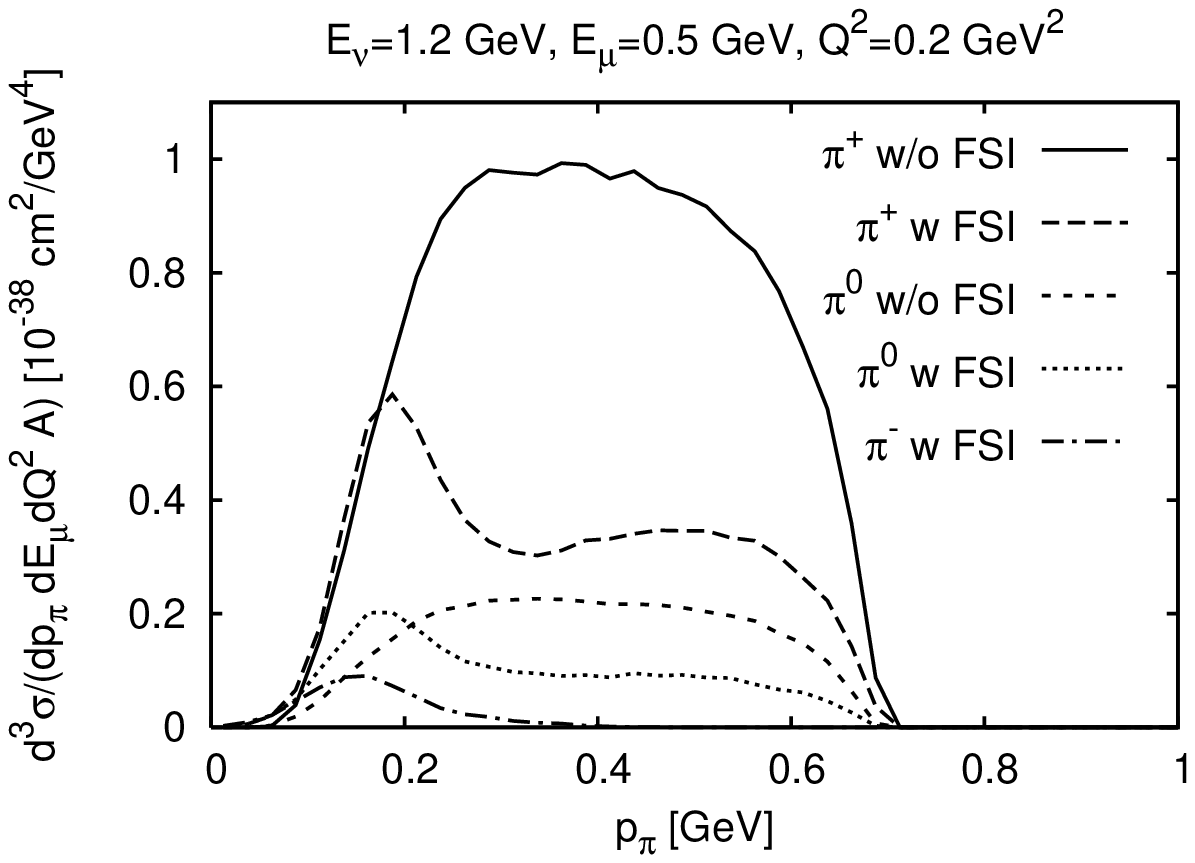,scale=.65}  
\caption{Momentum differential cross section for $\pi^+$, $\pi^0$ and $\pi^-$.  \label{fig:momentum}}  
\end{figure}

The presented model is applicable for CC and NC scattering of $\nu$ and $\bar{\nu}$ of all flavors off any nucleus. Our results for pion production and nucleon knockout are presented in detail in \cite{leitner}. Here we discuss exemplarily the results for CC pion production induced by $\nu_{\mu}$ on $^{56}$Fe in the energy range of $E_{\nu}=0.7 - 1.5$ GeV. Pion production is dominated by the initial $\Delta$ excitation; quasielastic scattering followed by pion production in $NN$ collisions is sizeable only at high $Q^2$. Fig.~\ref{fig:pionprod} shows that the total $\pi^+$ and $\pi^0$ production cross section is significantly reduced due to absorption when FSI are included. Note that $\pi^-$ are not produced in the initial $\nu N$ reaction but only through FSI. For $\pi^0$ production, we find an enhancement at low pion momentum $p_{\pi}$ due to the dominant $\pi^+$ production followed by charge exchange (Fig.~\ref{fig:momentum}). 

%\section{Conclusion}
We conclude that in-medium effects in $\nu A$ scattering, and in particular FSI, are not negligible. Their understanding - within a well tested transport model - is crucial for current and future experiments.

\end{document}